\documentclass{article}

\usepackage[letterpaper, textheight=9.0in, textwidth=5.2in]{geometry}
\usepackage{setspace}
\usepackage{fancyhdr}
\sloppypar
\newcommand{\documentname}{\textsl{white paper}}
\newcommand{\sectionname}{Section}
\newcommand{\secref}[1]{\sectionname~\ref{#1}}
\newcommand{\notename}{note}
\newcommand{\noteref}[1]{\notename\textsuperscript{\ref{#1}}}
\newcommand{\paragraphskip}{\medskip}
\renewcommand{\paragraph}[1]{\par\paragraphskip\noindent\textbf{#1}}
\let\OLDthebibliography\thebibliography
\renewcommand\thebibliography[1]{
  \OLDthebibliography{#1}\setlength{\parskip}{0pt}\setlength{\itemsep}{3pt plus 0.3ex}\footnotesize\raggedright%
}
\usepackage{url}

\begin{document}

\section*{\bfseries Why do we do astrophysics?}\thispagestyle{empty}
\noindent{\textit{by} \textbf{David W. Hogg}\footnote{%
The author is at the Center for Cosmology and Particle Physics, Department of Physics, New York University; and at the Center for Computational Astronomy, Flatiron Institute; and at the Max-Planck-Institut f\"ur Astronomie, Heidelberg.

I am thankful to
  Roger Blandford (Stanford),
  Mike Blanton (Carnegie),
  Josh Bloom (Berkeley),
  Gaby Contardo (Nova Gorica),
  Matt Daunt (NYU),
  Leslie Greengard (Flatiron),
  Juna Kollmeier (Carnegie),
  Jim Peebles (Princeton),
  Adrian Price-Whelan (Flatiron),
  Hans-Walter Rix (MPIA), and
  Kate Storey-Fisher (Stanford)
for discussions over the years, which informed the ideas presented here.
In addition I want to specifically thank
  Blanton,
  Andy Casey (Flatiron),
  Natalie~B. Hogg (Cambridge),
  Teresa Huang (Flatiron),
  Phil Marshall (SLAC),
  Greg McDonald (Rum\&Code),
  David~R. McMillen (Toronto),
  Gautham Narayan (UIUC),
  Monica Pate (NYU),
  Rix,
  Storey-Fisher,
  Soledad Villar (JHU),
  and an interactive audience at the SkAI Center in Chicago
for discussion of the particular content of this \documentname.
None of these people are responsible for any of the claims presented here.
The Flatiron Institute is a division of the Simons Foundation.
I acknowledge that the land politically designated as New York City is the homeland of the Lenape (Lenapehoking), who were unjustly displaced.}}

\paragraph{Abstract:}
At time of writing, large language models (LLMs) are beginning to obtain the ability to design, execute, write up, and referee scientific projects on the data-science side of astrophysics.
What implications does this have for our profession?
In this \documentname{},
I list---and argue for---a set of facts or ``points of agreement'' about what astrophysics is, or should be;
these include considerations of novelty, people-centrism, trust, and (the lack of) clinical value.
I then list and discuss every possible benefit that astrophysics can be seen as bringing to us, and to science, and to universities, and to the world;
these include considerations of love, weaponry, and personal (and personnel) development.
I conclude with a discussion of two possible (extreme and bad) policy recommendations related to the use of LLMs in astrophysics, dubbed ``let-them-cook'' and ``ban-and-punish.''
I argue strongly against both of these; it is not going to be easy to develop or adopt good moderate policies.

\section{Introduction}\label{sec:intro}
This document is being written at a moment of very rapid change.
What are currently called ``large language models'' (LLMs)---%
or sometimes ``generative artificial intelligence'' (although I really don't like that)---%
can now design, plan, execute, write up, and referee astrophysics projects.
At time of writing, one scientific-paper-by-LLM project has already been published \cite{denario} and I expect more in the category to follow.
There are already conferences being set up in which all the papers are written, and all the refereeing is performed, by machines \cite{conference}.
There is evidence that a large fraction of submitted academic papers contain LLM-written text and the fraction seems to be growing with time \cite{paperslop}.
This \documentname{} is inspired by this very rapid change.

For context, LLMs are models trained on huge text corpuses (close to the entirety of all human text output, ever) that can fill in and predict missing words in text or extend text starting at prompts from users.
These models produce informative, syntactically correct text, reproducing facts and ideas in the training data, but in new phrasings and with apparently synthetic recombinations of ideas from disparate sources.
These LLMs are capable of generating novel, interesting, and correct content.
I, personally, don't think they represent ``generative artificial intelligence,'' because although they are generative (they generate new things), they show no signs of intelligence.
They do, however, pass the Turing Test \cite{turing}, which is certainly concerning.

Unrelated to the rise of LLMs and their increasing academic capabilities, there is another change happening in astrophysics, which is happening on a much longer time scale, but which is relevant:
Astronomical data production is becoming extremely professionalized, and in a very particular way.
An extreme example is the ESA \textsl{Gaia} Mission \cite{gaia}:
Once the instrument was designed, the plans were moved inside the fence\footnote{I am using the phrase ``inside the fence'' to mean ``accessible only to those with governmental approval to work on weapons systems, that is, those with security clearances.''} at EADS, the European military contractor.
EADS then figured out how to build it, built it, launched it, and operated it.
No research-active astronomers were involved in any part of that operation; indeed even the scientific members of the \textsl{Gaia} Data Processing and Analysis Consortium (DPAC; \cite{dpac}) are in the position of having to reverse engineer some details of the on-board decisions (on the \textsl{Gaia} spacecraft) about what to window and telemeter to ground.
Don't get me wrong: \textsl{Gaia} has been incredibly important to global astrophysics (and my own research program); it is probably the most productive astrophysics space mission in science-per-dollar terms in the last decade; but it wasn't built or operated by \emph{astronomers}.\footnote{%
I will use the words ``astronomer'' and ``astrophysicist'' almost interchangeably here. For me the distinction is soft, and related to \emph{intentions}:
Astronomers want to measure things in the sky precisely. Astrophysicists want to derive physical insights from those measurements.
Most of us do a bit of both (I hope).
Both astronomers and astrophysicists contribute to and criticize the scientific literature. One could object to my claim that \textsl{Gaia} was not built by astronomers by saying that any employee of EADS who worked on \textsl{Gaia} is obviously an astronomer:
They built a telescope!
But they are not astronomers in the sense of contributing to and criticizing the astronomical literature.
Indeed, some employees at EADS are not even permitted to discuss their work in public, because their work relates to weapons systems.}
Astronomers, in the case of \textsl{Gaia}, are just end users; end users of curated, calibrated data, delivered by a combination of the (secret) spacecraft and the (absolutely great, professional, and open) DPAC.

The \textsl{Gaia} Mission is an extreme example, but many large projects are in this category, in one way or another:
Most of the NASA \textsl{JWST} \cite{jwst} instruments were built and delivered by defense contractors.
The enormous \textsl{LSST} project \cite{lsst} is being operated by a team that is a mixture of astrophysicists and engineering professionals.
Even university-based projects like the \textsl{Sloan Digital Sky Survey} projects \cite{sdssiv, sdssv} or \textsl{GALAH} \cite{galah} endeavor to produce science-ready data products that can be queried through application programming interfaces, plotted, and analyzed without much worry about where they came from or how they got here.
There are two points in play here:
The first is that teams that \emph{produce data} are increasingly professionalized, and removed from research astrophysics.
The second is that the astrophysics community is increasingly expecting to receive trustworthy, vetted, calibrated, and complete data.

I have been involved in bringing about this change (for example, \cite{monitor, ubercal, astrometrynet, opendata}), and in many ways it is absolutely great.
It democratizes astronomy, since it lowers barriers to entry.
It creates an open-science space, in which every project benefits from the output of every other project.
It makes it very easy to make discoveries, and especially discoveries that involve multiple data modalities.
It thus weakened the walls of the astronomical silos (``radio astronomer,'' ``spectroscopist,'' and so on), and especially blurred the distinctions between the categories of ``theorist'' and ``observer.''
But it does have a strange consequence, which is also related to professionalization:
For some kinds of projects in astrophysics, there isn't a huge difference in capability between a classically trained astronomer and a newly trained \emph{data scientist}.
Indeed, in many ways, the data scientist is better prepared to work in astronomy than the astronomer, since the data scientist knows how to do fast computation on large data sets, build complex probabilistic models, and perform large-scale optimizations or searches.
For analysis of the \textsl{Gaia} data, a data scientist who has taken an astronomy class might be better prepared than an astronomer who has taken a data science class.

Of course much of astrophysics is \emph{not} like working with the \textsl{Gaia} data.
Many astrophysicists build instruments, operate observatories, and theorize.
Even a student observing at a small, university-based observatory is doing things that go way beyond data science, in assessing conditions, deciding things about data quality, or making on-the-fly changes to the observing schedule based on weather or issues at the telescope.
Not all of astrophysics is data science, not by a long shot!
But a lot of it is, and the fraction is increasing every year.

How are these two changes---LLM successes and the data-scientification---related?
They are related because \emph{contemporary LLMs can do data science}.
It's true that there are many known limitations of present-day LLMs, for example in doing basic mathematics \cite{arithmetic}, or in performing causal inferences \cite{causal}, but the models are improving by the week, and I think it is reasonable to expect that they will steadily become more competent at designing, executing, writing up, and criticizing scientific projects, at least for a while.
There might be show-stoppers; we don't know yet.
But this \documentname{} is being written under the assumption that LLM-like models (or ``genAIs'') will improve with time until they can perform data science tasks with the same proficiency as humans.
What does the study of astrophysics look like in this future?
I don't know; this \documentname{} isn't even going to answer that.
All I am going to ask is: Why do we do astrophysics?
I have the strong feeling that we will need to answer this question first, before we can answer questions about our relationships with LLMs and LLM-based projects.

\section{Points of agreement}\label{sec:facts}
We'll start here by stating some ``facts'' about astrophysics, or points on which we can all agree; ``points of agreement'' in what follows.
Of course for each of these, there are astrophysicists who will disagree, because there are as many ideas of what astrophysics is as there are astrophysicists.
But these form a set of assumptions from which I will try to flow my criticisms of policies or practices in \secref{sec:policy}.

\paragraph{Astrophysics produces new knowledge about the Universe.}
Astrophysics is a science that is driven by discovery and new measurement.
Every PhD project and (almost) every paper involves measuring something that hasn't been measured before, or in a different way, or making a new interpretation or prediction, or improving a methodology, or finding a new object or set of objects.
That is, every project that counts as ``astrophysics'' involves scientific novelty.

If you want to become an astrophysicist, it isn't sufficient to read about it, or take classes in it.
You have to \emph{do it}, and \emph{doing it} requires doing novel things, that haven't been done before, and which connect to important scientific questions in the literature.
It also involves (as I will discuss below) writing up and publishing what you did.

It isn't even sufficient to repeat classic or known measurements.
Part of becoming an astrophysicist is learning how to design and plan new kinds of measurements or new approaches or applying known techniques in new areas to make new discoveries.
It also involves situating, in writing, that novel work within the contemporary scientific literature.
The novelty of the work is explicitly part of the project; astrophysics isn't just techniques or results; it intrinsically involves innovation.
Astrophysics lives at the cutting edge,\footnote{%
I owe a lot to Rosie Wyse (JHU), in general, but also in particular for making this point very clearly to me in a conversation in 2023.}
not in the past.

\paragraph{People are always the ends, not merely the means.}
It might sound trite to say that we do astrophysics for people.
But the consequences of this are at least a little bit non-trivial:
When we employ a graduate student to perform some work, it absolutely must be \emph{because the graduate student will benefit} from that work, not merely because that work needs to get done.
This is one of the interpretations of the categorical imperative \cite{kant}, which is (to my mind) one of the few undeniably true principles or laws that have ever been written in the philosophical subject of ethics.

I have heard it said, more than once, in research contexts, that an LLM can do some task ``better than a graduate student.''
That language makes me uncomfortable, because it is taking an extremely \emph{instrumental} view of graduate students.\footnote{%
By far the most generous interpretation we can give to a comment involving the phrase ``better than a graduate student'' is something like what we mean when we say that a calculator does long division ``better than a seventh grader.''
Seventh graders shouldn't be doing long division (except maybe to originally learn the concept).
But I don't think this interpretation is correct, because the tasks about which this is said are tasks which, unlike seventh graders and long division, PhD students in astrophysics \emph{very much should be doing}, like writing code, reading literature, and visualizing results.}
Are graduate students in our groups and our laboratories and our universities \emph{to do work}?
Or are they here \emph{to learn}?
I very much hope the latter, or, if they are here to work, it is because that work is also critical to their learning.
We train PhD students not merely to amplify our own research programs, but to create opportunities, and specifically opportunities for them.
In general, it is possible to take a very instrumental view of the world, in which people are described in terms of their functions---their input--output relationships, if you will---and this is the view of the world I often see reflected in the press releases of robotics and artificial-intelligence companies.
It is important that we reject this view.
Every person is a human being, whose personal development is more important than our short-term scientific accomplishments.\footnote{%
One could make an argument that our \emph{long-term} scientific accomplishments---the theory of gravity, the discovery of dark matter, the possible future understanding of the origin of life or the discovery of life outside the Solar System---might be as important as people, in some sense.
But individual things that PhD students are doing are way less important than those PhD students themselves.}

You might object: But science is \emph{also} important!
It is, perhaps, but see below my point about clinical value.
If a science does not have clinical value (no way to help people), then its short-term results are definitely and absolutely less important than its people.
If a science \emph{does} have clinical value, there might be room for debate.
You might notice that I am not a \emph{utilitarian}, but I am willing to listen to utilitarian arguments.\footnote{%
Utilitarian arguments balance benefits and harms to people, with the goal of finding the solutions that most increase the ``sum total of human happiness'' \cite{mill}.
If a science has clinical value, then the harm to workers (of being forced to do things they don't want to do, say, or being underpaid for their work, say), can be balanced by the good done to the recipients of the clinical successes.
I don't like these arguments, because they are used to justify using people as a means, and not an end.
Utilitarianism works well with capitalism and development.}

One interesting possible consequence of this point---the point that people are the ends, not the means---is that the standard form for research proposals might be unethical:
Research funding in those proposals pays for graduate students, with the explicit promise of deliverables from those students.
When it's crunch time, are we obeying the categorical imperative?
Another interesting possible consequence is that \emph{scientific embargoes might be unethical}.
After all, how does stopping early-career scientists from talking publicly about their work serve \emph{their} interests?
I have a strong feeling that there are also important consequences here for the structure of qualifying exams and interviews and other gates in our educational, hiring, and promotion systems, which are all laden in old traditions.

Much of what I have written here is about students, but it applies more broadly.
Indeed it applies to everyone in the scientific enterprise,
including all researchers, from junior to senior.
Every scientific paper is written to help all of its writers, and all of its readers, learn and grow, no matter their career stages.
Thinking \emph{even more broadly}, when I say that ``people'' are always the ends, I don't mean to be human-centric.
If we started doing astrophysics with deep-ocean octopuses,
or extra-terrestrial intelligences,
or truly intelligent machines, I think it is probably the case that those other intelligent entities probably also would have to be the ends, and not merely the means of our work.\footnote{%
I am not certain that it is \emph{intelligence}, precisely, that makes an entity deserving of inclusion in the categorical imperative, but whatever it is, I doubt it is ``membership in the human species'' \cite{animallib}.}

\paragraph{Astrophysics is (roughly) the astrophysics literature.}
My group is, in some sense, a software group.
We produce software systems that many people use (for example, \cite{astrometrynet, emcee}).
For this reason, I am occasionally asked to speak on the idea that software is just as important as other kinds of scientific outputs.
People are surprised when I don't completely agree.
I believe that (in astrophysics) software is written to support the astrophysics literature, and that every important piece of software should have an associated paper in the astrophysics literature.
That paper describes the intellectual contributions encoded in the software, and provides a standardized mechanism for tracing intellectual provenance, giving credit, and coordinating criticism.
Maybe my position is radical now?
But I believe that software is important in that it embodies and executes \emph{our ideas and beliefs} about astrophysics.
These ideas and beliefs are well gathered, disseminated, and preserved in the astrophysics literature.\footnote{%
I should say here that I take the dissemination and preservation aspects of the literature more seriously than any other aspects.
For this reason, I believe that \textsl{arXiv} is every bit as important as \textit{The Astrophysical Journal}.
However, I believe that low-cost ``free journals'' that just provide a ``refereeing layer'' don't contribute much;
actually preserving and disseminating knowledge is never inexpensive.}

Nothing we have has the track record that the traditional journals do for long-term dissemination and preservation of knowledge; we don't know what computing hardware platforms we will have in 15 years, let alone 200, but I can pretty-much guarantee that---if human civilization is still in existence---we will still be able to read Vera Rubin's papers about the dark matter (for example, \cite{vera}).\footnote{%
A seriously off-topic comment is that with the the journals no longer having print (hard-copy) form, we are one small-scale war away from losing everything that astronomers have done this entire century.
That's why my example paper is Vera Rubin's, not any of my students'.}

My point is that astrophysics \emph{is} the astrophysics literature.
If you want to ask ``what is known about the structure of neutron stars?'' you ask the literature (scientific articles plus sometimes textbook chapters).
Even if you ask a person, that person justifies their answer by pointing to the literature.
If an astronomer measures something, but never publishes that measurement, then that measurement does not get disseminated, known, and preserved.
It does not become part of astrophysics.
That is, the literature is the repository of all of our knowledge and the only reliable authority.

I will make a few small comments about this.
The first is that astrophysics, like any science, contains a lot of ``implicit knowledge'' or folklore about things like how to observe, how to reduce data, how to organize projects, how to visualize data and models, how to read and write, and so on.
Much of this never appears in the literature.
Is that not also astrophysics?
Yes it is, but it is \emph{astrophysics practice}.\footnote{I would welcome a project in which we tried to make much of this implicit knowledge explicit.}
The results of astrophysics---the scientific conclusions and debates---are in the literature.

The second comment is that I often hear software (and hardware and engineering-oriented) people say that they ``have to'' write papers because papers---and the citations that they generate---are ``the coin of the realm.''\footnote{%
I have heard C.~Titus Brown (UC Davis) present the requirements of traditional publishing as being at least marginally in conflict with open science and scientific open-source software development.}
Papers (and the authorships on those papers) and the citations of those papers are not ``coin'' of anything!
They represent our recording of what happened, what we learned, what we know, and how we know it.
Citations deliver provenance, not reward.\footnote{%
The authors on a paper should be those responsible for creating the results described in that paper.
Papers should cite the papers that are relevant to their results---the papers they are criticizing or building on.
Anything else unethically distorts the provenance of the ideas and work in our field.}
All that said, and as will come up again in \secref{sec:policy}, it is true that publication counts and citation counts are heavily used in hiring, promotion, and grant-award decisions; citations may, effectively, be the coin of the realm even if I don't want them to be.
One comment to make about authorship is that it is probably not the case that an LLM can be the author of a scientific paper.
That is, it isn't really correct, if you use an LLM, to give that LLM co-authorship on your paper, any more than it is to give \textsl{astropy} \cite{astropy} or \textsl{GitHub copilot} or the \textsl{Mac} operating system coauthorship.
Authors, because they are taking responsibility for their work, need to be the kinds of entities that can take responsibility for their work.
Right now, LLMs can't take responsibility for anything, and they can't, later on, answer questions about why or how they did what they did (well they can, but not because they remembered it, only by reverse-engineering it).\footnote{%
Things might get weirdly different if LLMs obtained very long-term memories (which isn't obviously difficult to implement), but it still probably wouldn't address the point that, in their current form, none of the LLMs has any way to \emph{take responsibility} for anything.}
Weirdly, that hasn't stopped the publication of some scattered papers with ``ChatGPT'' as a coauthor, for example \cite{chatauthor}.

One way to codify this responsibility point has been put forward by the Committee on Publication Ethics \cite{cope} and the International Committee of Medical Journal Editors \cite{icmje}.
One of the reasons that an LLM cannot be an author because it \emph{cannot sign a copyright or licensing agreement}.
That argument is too connected to technical intellectual property rules for my taste, but it does make legally explicit the point that LLMs cannot take responsibility for anything.

Finally, because the astrophysics literature preserves provenance, it is ethically required that our papers cite the work that is relevant to the work we are doing.
You can't decide not to cite a relevant paper because you don't like the author, or don't like the author's institution, or don't like their funding sources.\footnote{%
This sentence is perhaps a bit strong.
There are extreme cases where very, very bad (exceedingly unethical, say) research practices render papers unciteable.
But these are rare in astrophysics.}
In particular, if the literature gets flooded with work of relevance to your research program, you have reading to do, and citing to do.
This connects also to the points of agreement here about novelty and efficiency and rigor.

\paragraph{Astrophysics (like every science) requires trust:}
You can't do science if you don't live within a network of trust.
This point is extremely general: You have to trust your coauthors, you have to trust the literature, and you have to trust the machinery and tools that you use.\footnote{%
This point has been made strongly to me many times by Mike Blanton (Carnegie), who gives a lot of credit to Oppenheimer \cite{oppy}.}
Obviously a critical part of science is testing and verifying the tools and the literature and the work we do.
Every paper we write does some partial verification of different parts of the scientific environment in which we work; we develop our trust as we work, and deliver effective verifications of different components to ourselves and others.
But it is impossible for an investigator to test and verify everything; the untested things must be trusted.
It would be very difficult to do astrophysics in a context in which there is a large part of the literature that we don't trust:
We would be required to cite it---citing relevant work is ethically required---but you can't cite things you don't trust.
Thus we would have to do an enormous (and possibly unsustainable) amount of verification and testing in that environment.

For the same reason that we must trust many entities in order to do our work,
it is incumbent on us to \emph{build trust}.
Open science, reproducibility, and collegiality all do this, and they all serve astrophysics well.

\paragraph{We must use our resources efficiently.}
Partly because we deliver no clinical value (see below), essentially all astrophysics grants and funding are \emph{gifts}.
Our funding comes from a combination of public purses (national agencies, public universities, national and international observatories and spacecraft) and private sources (foundations, private universities, individual donors).
I don't understand the motivations of most of these actors, if they can even be described as having motivations at all.
However, if we squander these funds, the sources will dry up.
Perhaps more importantly, when the source of our funding or support is public, we owe it to the people represented by that public entity to use their contributions wisely and well.
Even with private funding, the source is, ultimately, the people.

It is an absolute requirement that we not waste money.
Therefore, if there are a few different ways to make the same measurement, it is incumbent on us to choose the most efficient of them, unless there is some important intellectual value in making the measurement in multiple ways (as there sometimes is).
We should work with the wavelengths, the techniques, the people, the technologies, and instruments that are best at making our measurements.

This point is connected to the ``learn new things'' point above, but it is worth noting that it is slightly in conflict with the ``people are the ends'' point.
After all, graduate students aren't \emph{labor} to be \emph{efficiently deployed}.\footnote{%
The first person to clearly say this to me was Mike Blanton (Carnegie), during a conversation at NYU about the efficiency of the \textit{Sloan Digital Sky Survey IV}.}
Graduate students are human beings, for whom we are supposed to be creating new opportunities.
Importantly for what follows, however, if LLMs can do something important, efficiently, it is not obvious that we can just ignore that.

\paragraph{Correctness and rigor are paramount.}
Astrophysics is an observational science.
We like to talk about ``experiments,'' but fundamentally we can't perform controlled experiments:
We observe the Universe as it is, with the photons that happen to enter the apertures of our telescopes during the times our shutters happen to be open.
Thus exact reproducibility is difficult, and some observations cannot be verified, no matter what we do.\footnote{%
For example, the very smallest, longest-period planets we found with the NASA \textsl{Kepler} Mission \cite{keplerresults} represent very low-amplitude, very rare photometric events.
They are essentially impossible to confirm with any independent data, short of launching another \textsl{Kepler}-like mission.
If that isn't extreme enough, consider supernovae:
No supernova ever repeats, so it would literally require a violation of Lorentz invariance to verify any observations, light echos \cite{lightecho} notwithstanding.}
Thus we check or ``confirm'' many measurements in astrophysics primarily or partially \emph{by inspection}, inspection of the data and code that were taken and used in the generating the analysis and conclusions.

The only authority we have for such confirmation lies in methodological domains adjacent to astronomy:
We rely on optics, mechanical engineering, remote sensing, statistics, and---very importantly---applied mathematics.
If we don't use these methods correctly, we will get wrong answers, and, in the daisy chain of reasoning that is astrophysics, those wrong answers will propagate through to other, downstream results.
It is imperative that we do our work as correctly as possible, and that we understand and admit the approximations and mistakes that we are making along the way.
We must test those assumptions and correct those mistakes, perhaps not immediately, but over time, and as a community.
We must call out mistakes and criticize wrong results and bad assumptions, and we must re-do analyses that we believe to be flawed.
If we stop doing things rigorously and correctly, we have stopped doing astrophysics.
This is especially true in a science with \emph{no clinical value}:
If our results don't have direct application to the real world, that is, applications which effectively test our results, then shoddy or incorrect results can live on and lead to more wrong results downstream.

\paragraph{We don't do astrophysics to \emph{find out the answers}:}
This is probably the least agreed-upon point of agreement, and my hottest take.
I claim that (essentially) none of us, individually, does astrophysics because we want to learn the specific answer to the astrophysics question we are asking.
\emph{What?}
For example, when we measure the age of the Universe, no-one on the team actually cares what the specific value is, even though each of them might have spent many hard years of their life figuring out how to make the measurement correctly (and it isn't easy).

This point---that we don't care about the specific results---is obvious, in the following way:
Anyone who has the capability of getting a PhD in astrophysics has the capability of doing many remunerative things, substantially more remunerative than my job.
Thus, anyone who is doing astrophysics (professionally) could be, instead, earning enough money to pay \emph{multiple} professional astrophysicists to work on their behalf.
That employed astrophysics team would get to the answers of any astrophysics questions faster than the individual could by working on their own.
If all we \emph{really} wanted was to know how the Universe worked, we would start a hedge fund, and use the proceeds to pay an astrophysics institute, filled with people who wanted to \emph{do astrophysics} rather than \emph{find out the answers}.
This is not far from what people like Jim Simons (co-founder of the Simons Foundation) did.
Jim Simons is partially responsible for a lot of results in physics and astronomy (and mathematics and Autism research, and more), more than any astrophysicist,
and he did all that by running an extremely successful hedge fund in the US \cite{simons}.
Thus anyone working in astrophysics is someone who wants to \emph{do astrophysics}, not someone who wants to \emph{learn the answers}; there are way more efficient ways to learn the answers.\footnote{%
This argument assumes that astrophysicists are, by and large, rational.
That's a bad assumption, given the points about remuneration.
However, I think the point is still good.
Also, I must comment that this argument is unfortunately related to an absurdity known in the tech sector as ``effective altruism'' \cite{effectivealtruism}.
For this I apologize; I think that effective altruism is wrong because of its assumptions about predictability of impacts, effects, and needs over long time scales.
(It also assumes that the evil of rising disparity can be overcome with individual good works, but that's even \emph{more} off-topic.)
The argument presented here does not make assumptions about the predictability of medium-term or long-term futures.}
This all would be different, I believe, if astrophysics had clinical value:
Then we would care about the detailed results, because, if they came out well, they would lead to new opportunities for humanity.

\paragraph{Astrophysics has no clinical value.}
I like to say that the sciences have a ``left edge'' which is about fundamental understanding, and understanding for understanding's sake.
They also mostly have a ``right edge'' which is about what I like to call ``clinical value'' but you could call application or use in the world for technologies or policies.
So, for example, biology has a left edge on which people understand how the cytoskeleton moves the cell, and a right edge in which they develop or improve therapies for cancer or cloudy toenails.
Macroeconomics has a left edge that involves relationships between unemployment and inflation, and a right edge that informs government policy about interest rates.
I claim (and maybe this is a bit controversial) that \emph{astronomy has no right edge}.
That is, there are no useful things in the world that flow from astronomical discoveries and results.
I have spent years of my life estimating the comoving volume of the Universe \cite{phdthesis}, measuring the local dark-matter density \cite{oti}, and finding planets around other stars \cite{singletransit}.
No human outcome or pragmatic capability has been affected in the slightest by any of my results.
Literally nothing helpful to humanity arises here.

I speak about this sometimes, and I hear objections of various kinds.
One silly objection is that we might find ``Planet B.''
Nope. We're all dying---and every human who is ever born will be dying---on this rock, or extremely close to it.\footnote{There are many reasons this is true; some of them are discussed in \cite{marssux}. It is also very eloquently stated by Michel Mayor in his acceptance speech for the 2019 Nobel Prize in Physics.}
A less silly objection is that the results of astrophysics inform particle physics.
They do!
We learned about neutrino mass mixing and the entire dark sector from astrophysics, not particle experiments.
But these astrophysics results inform the \emph{left edge} of particle physics, not the right edge.
An even less silly objection is that we develop, build, and exercise imagers, cameras, spectrographs, spacecraft platforms and so on, which have right-edge applications.
All that is true!
That---with the exception of the Earth-crossing asteroids, mentioned above---is as close as we come to having a right edge:
We contribute to precision remote-sensing technologies, which, in turn, help the US and other nations target weapons.
Congratulations, astronomy!
But I don't even count this as clinical use of astronomy, because we build these tools \emph{to do astronomy}, these aren't the \emph{results of astronomy}.
My point about astronomy having no right edge is that nothing in the world of things or people hangs on the precise value of the age of the Universe.
Many astronomers have spent many years (and billions of dollars) measuring it, and we know it very precisely \cite{age}.
But literally nothing hinges on the question of whether it is 13.77 billion years or 13.79.
So feel free to disagree here, and say ``yes, astronomy does have a right edge; we improve the targeting of weapons.''
I won't fight you.
But, for the purposes of my arguments, this won't be the kind of right edge that affects how we do things.

One way to see this is that even if our detectors and satellites were useless to any other industry,
we wouldn't use them differently, or do anything differently.
Note the difference with biology: If we found that cancer could not ever be cured with biological resources, a large part of the study of biology would change overnight.
After all, much of biology is justified (albeit sometimes extremely performatively) in these right-edge terms.
None of astrophysics is justified in these right-edge terms.
No astronomer (that I know) is improving the calibration of \textsl{JWST} instruments because they want the US Navy to have a higher kill rate.\footnote{%
The closest thing I can see in my own world to anyone making this argument is that the US funding agencies (NASA and NSF) explicitly prefer research proposals that develop technologies that could be transferred to commercial enterprises.}

Another way to say all this is that astrophysics may occasionally and accidentally produce something useful.
But astrophysics is not done (any more, in Western contexts\footnote{%
In non-Western and ancient contexts, astronomy did have important use value in societies.
I'll return to that point at the very end of \secref{sec:policy}.})
with the \emph{goal} of obtaining clinical or practical value.
Clinical value is not part of the \emph{intent} or \emph{goal} of the work in astrophysics.\footnote{This teleological argument was brought to me by Andy Casey (Flatiron).}

While I want to say, in this \sectionname, ``no clinical value,''
I have to admit that there is one substantial clinical value of one sub-branch of astrophysics:
The search for Earth-crossing asteroids might potentially, some day, save humanity from an enormous disaster.
I think this is niche enough to leave my main point untouched,
but I do have to admit that the work we do on instrument calibration and the detection of faint, moving objects (my own contribution here being \cite{undetectable}) could deliver enormous, direct benefits for humanity some day.\footnote{%
Kate Storey-Fisher (Stanford) pointed this out to me and also quipped that it probably isn't really a clinical value, however,
since if we discover an asteroid on an Earth-collision trajectory we will probably all just ignore it anyway; see \cite{dontlook}.}

When I make the no-clinical-value argument I sometimes get the following:
Since (as we said above) humanity loves astrophysics, it must be the case that it is clinically valuable!
It must be that the clinical value lies in its feeding of humanity's love.\footnote{%
Roger Blandford (Stanford) sometimes says that astrophysics provides ``hope'' to humanity.
Maybe that's a strong claim, but I don't think it is wrong.}
But this is not truly a \emph{clinical} value in the following sense:
In no way does this application of astrophysics depend on the correctness or values of the outcomes of astrophysics projects.
For example, if biology learns something new about a protein interaction, that might directly affect a drug-development path.
If that result turns out to be wrong, that drug-development path will fail.
The clinical side of biology \emph{tests} biological results.
Similarly, if the macroeconomics community gets the relationship between unemployment and inflation wrong, policy adjustments by central banks will fail to work.
A science has a right edge if and only if the associated clinical work actually tests or exercises the specific results of the science.
In this sense, astrophysics has no right edge.

\section{The benefits of doing astrophysics}\label{sec:benefits}
\secref{sec:facts} was supposed to list some facts, or points of agreement, about what astrophysics is,
or what it should be.
One of those facts is that astrophysics has no clinical value---its results provide no useful resources to humanity.
That doesn't mean there aren't extremely important benefits of doing astrophysics.
This \sectionname{} is about those benefits.
What do we deliver to the world?
I am going to list all the things I know here, but (as you will see) I don't agree with all of them.
When I say ``all the things I know,'' I really am attempting to be complete:
I am trying to list here any benefit that any entity (person or organization) could feel or receive from any part (effort or result) in astrophysics.

\paragraph{People love astronomy.}
When I decided to work in astrophysics (which I did in the period 1990--1993), I remember my father saying that he felt like the area is a good one, because astronomy will always be studied by humanity, no matter what.
The interest in astronomy is so deep, relevant research has been done by countless civilizations, as long as humans have been around (and probably before).
In the same way that it is good that humans produce (and read) novels, and poetry, it is good that humans produce (and learn) new ideas in astronomy and astrophysics.

Related to this, astronomy inspires and (almost) produces art.
Many beautiful paintings, photographs, and sculptures have been inspired by astronomy, and music has been composed and performed in honor of the heavens.
Whole genres of science fiction are directly inspired by discoveries in astronomy and astrophysics,
and, correspondingly, some astrophysics can even be learned from reading ``hard'' science fiction.

\paragraph{We create human knowledge.}
One of the tremendous accomplishments of humanity is its scientific knowledge, represented by the scientific literature, and also many other non-western records of science.
It is obviously a good thing that humanity has created, preserved, and transmitted this knowledge across space and time.
Thus our contributions to the astrophysics literature are obviously good:
They are contributions to a very good thing.

\paragraph{Universities need astrophysicists.}
We put ``the universe'' in ``the university''?
A university is a place where students can study and learn \emph{anything} of importance to humanity.
Astrophysics is important to humanity.
Therefore it is essential that universities have astrophysicists in faculty positions.
Inasmuch as the research university is a good idea, the training and hiring of astrophysicists, and the study of astrophysics, are all good ideas.

\paragraph{We train a technical workforce.}
Very few of my PhD students---and only a small fraction of all astrophysics PhD students, globally---become research astrophysicists.
Most astrophysics PhD recipients obtain long-term careers in other technical professions, such as tech, finance, and engineering.
Astrophysics, it turns out, is a very good training for quantitative professions that involve computational modeling.
Thus astrophysics research produces very good people for technical innovation and growing our economy.
This technical workforce argument is literally and explicitly the reason that the Max Planck Society in Germany supports astrophysics research.\footnote{Hans-Walter Rix (MPIA), private communication.}

Although PhD students are the group most ``trained in astrophysics,'' all the same arguments apply to undergraduates.
Undergraduate astrophysics students go on to technical careers in all kinds of industries,
and also populate graduate schools in many natural-science and engineering disciplines.
An astrophysics undergraduate degree is a great preparation for work in climate, for example, which is undeniably important.\footnote{%
I've heard it said that a benefit of astrophysics is that, by employing astrophysicists, we keep strong technical people \emph{out of} other technical areas that are more harmful, like finance or weapons.
I don't agree with this!}

\paragraph{We educate the public.}
The public---from undergraduates \emph{not} in the sciences to the television-watching and TikTok-swiping publics---are excited about astrophysics and want to learn more about it and more from it.
I don't think it is an exaggeration to say that most members of society are interested in the beginning of the Universe, the properties of black holes, and the origin of life.
Heck, we might even discover aliens some day.
If the interests of the public matter, then we are doing something important to many, and contributing rich content to public education.
One thing many US astrophysicists like to mention is that NASA regularly gets coverage on the front pages of the newspapers of record (or I guess homepages, nowadays).
Every one of those instances is a huge success of research astrophysics, and contributes to public understanding of (and excitement about) science.

\paragraph{Physics needs astrophysics.}
As mentioned above, there are many things we can't understand about physics without doing astrophysics.
The highest vacuums, strongest electromagnetic fields, and largest length scales are only accessible astronomically.
The theory of gravity was found and is entirely informed (except perhaps in studies like \cite{adelberger}) through astrophysics, including black holes, gravitational radiation, and accretion flows.
The dark matter and vacuum energy density---and whole dark sector, whatever it is---is detectable (so far) only through stellar and cosmological dynamics.
Those are all aspects of fundamental physics that are available through astrophysics, and only astrophysics.

There are other physics questions that are less fundamental (perhaps), related to our \emph{origins}:
How did life form? How did the Earth form? How did the Sun form?
These are critical physics questions, and they also depend crucially on astronomical observations and astrophysical interpretation.

\paragraph{We beat ploughshares into swords.}
The budgets of Western nations\footnote{%
I don't want to exclude India, China, and Japan from this discussion, because I believe that the same things happen in Asia as well.
I just can't speak about it as confidently as I can about Europe and North America.}
(and especially the US) involve enormous contributions to military contractors, who in turn deliver extremely sophisticated technical capabilities for western-controlled weapons.
Despite this, there is a big and well-funded contractor lobby plus many lawmakers who believe that \emph{not enough} money is being delivered to these contractors.
Astrophysics projects create opportunities to use civilian budget (scientific research budget) on military applications.
If you want examples, look at the NASA and ESA roadmaps, which will show flying and near-future spacecraft, essentially every one of which delivers between hundreds of millions and billions of USD to weapons contractors.
This argument may make the reader uncomfortable---I sure don't like it---but this is absolutely one of the ``benefits'' of astrophysics, when one takes the view of ``national priorities.''\footnote{%
This ploughshares-to-swords activity is usually in play when boosters of NASA missions argue that, if we don't build and launch some particular mission, we will ``lose our national leadership'' in some technical area.
Those arguments are bad, I think, because they are primarily military arguments, and, also, they go against the upcoming ``borderless world'' argument.}

\paragraph{We use and develop remote-sensing capabilities.}
The astrometric and photometric precision requirements of contemporary astrophysics projects exceed those of remote-sensing projects in almost any other domain.
For this reason, astrophysics missions have been invaluable in the development of extremely stable (in an angular and thermal sense) spacecraft platforms, and of empirical models of the response functions of infrared detectors, for two examples of remote-sensing technologies.
We also exercise and contribute to the deep-space network, on-board compression and data management, and other engineering capabilities that assist with making complex measurements in low-bandwidth environments.
Many of the capabilities developed here are military, but not all are.
For example, ocean-observing missions \cite{oceanobserving} and precision tests of gravity \cite{gravityprobeb} have made use of technologies that were at least partially developed in astrophysics contexts.

\paragraph{We create opportunities for development.}
Despite the large budgets for some astronomical projects, we don't often think of ourselves as ``land developers.''
However, large projects on mountains in the US (including Hawai{\textquotesingle}i), China, and Chile (for examples), all represented development opportunities that contributed to economies (and often upset local populations).
In building they employ contractors and in operations they employ long-term staff; they make use of local utilities and services.
In some cases they improve local services or bring services to under-served locations.
In addition to these tangible effects, astronomical projects sometimes also effectively become monuments, representing national or regional pride (on the one hand, and colonial or state power on the other).

\paragraph{Astrophysics represents a borderless world.}
I have spent a large fraction of my career working on different generations of the \textsl{Sloan Digital Sky Survey} \cite{sdss, sdssiii, sdssiv, sdssv}.
These projects have become large, international collaborations with partners all over the Globe.
In the recent generations, institutions were invited to become partners in these projects without regard to national or geographic location, and indeed there are partners on multiple continents.
Similarly, when astrophysicists consider applicants for PhD programs, postdoctoral positions, and faculty jobs, we are open to applications from all over the world (although negatively constrained by visa or immigration requirements).\footnote{%
This is in stark contrast, by the way, to the scientific practice when I was an undergraduate.
The Cold War and the Iron Curtain split the world into (at least) two disjoint pieces, even scientifically.
I met very few of my peers in China or Eastern Europe prior to 1992-ish.}

It isn't just about institutions and people, it is also the literature and ideas:
The journals accept paper submissions without regard to country of origin, and have the same refereeing standards for contributions from any place.
There are obviously national-origin biases among editors and referees, and the literature is in English, and there complexities around page charges, but there are no explicit rules about country of origin for our principal journals.
Even more importantly, we don't (or certainly aren't supposed to) judge the correctness of a scientific claim based on the national origin of the scientist making the claim.

In all of these ways, astrophysics represents a borderless world, in which we demonstrate that borders are not required for the accomplishment of great things.
Just the opposite: The raising of walls is detrimental to science, as it impairs the network of collaboration, conferences, and the literature, which in turn compromises the mechanisms by which scientific ideas are tested, criticized, and elaborated.
Borders hinder, not help, the most important human activities and achievements.

\paragraph{Astrophysics is a satisfying activity.}
Finally, of course, \emph{astrophysics is fun}.
Most of us get involved because we are curious and like to solve problems and to build and do things.
The discovery of an object, the making of a measurement, the execution of a pipeline, the submission of a manuscript are all exciting and rewarding in themselves, and help us grow.
It is exciting to be a part of the community that figures out how the Universe works.
2019 Nobel Laureate Jim Peebles often speaks of the great success of ``curiosity-driven research'' in creating the empirical and theoretical bases for the physical model of cosmology that is ascendent today \cite{curiosity}.
I agree!

\section{Policy non-recommendations}\label{sec:policy}
I opened with the challenge confronting astrophysics with the arrival of increasingly competent large language models.
As I said, I won't end up actually making any policy recommendations here; this \documentname{} is a gathering of relevant ideas.
I will, however, suggest two possible, extreme policies; and I will argue \emph{against} both of them.
That is, I only have negative things to say about simple policies---or at least about extreme policies.

\paragraph{Let-them-cook:}
One possible extreme policy proposal I could make (and, to be clear, I am \emph{not} making) is that we \emph{fully embrace} the LLMs, and encourage them (or even give them enormous resources\footnote{%
One thing I have not addressed at all in this \documentname{} is the tremendous environmental impact of the training of LLMs, and their inferences (their execution as generative models working on prompts).
The numbers are staggering.
Of course it is important to remember that the human practice of astrophysics, at least in its current form, is also very damaging to the environment \cite{low-carbon, imperative, water}.\label{note:environment}})
to conceive of, design, execute, write up, and submit (to the journals) novel astrophysics projects and papers.
Under this policy, the vast, vast majority of astrophysics would be done by machines, and we (as a community of humans) would participate primarily through reading, vetting, and discussing the novel astrophysical results produced by those machines.
Dear reader, you probably either think this policy is disastrously wrong, or else obviously right.
I want to take it seriously, not least because it is the literal, explicit proposal of one recent project \cite{denario}, which has built a first version of the key tools necessary to implement this policy, end-to-end.

If we implement the let-them-cook policy (that is, full acceptance), astrophysics changes dramatically, but it is consistent with some of our points of agreement (listed in \secref{sec:facts}), and it does deliver many of the key benefits of astrophysics (listed in \secref{sec:benefits}).
On the points of agreement: Astrophysics-by-LLM produces new knowledge; it doesn't use people (therefore it doesn't make people its ends); it contributes to the literature; it is efficient (time-efficient---it does, after all, use energy resources; see \noteref{note:environment}).
It is not clear that it will be rigorous and correct.
Right now it is hard to assess the correctness of LLM-generated content, or at least it is not easier to figure that out for LLM-generated content than any other kinds of refereeing.
At the rate at which LLMs can design and produce papers, we humans probably can't keep control of rigor and correctness.
Aside from that, astrophysics by LLM is indeed properly astrophysics, according to most of our points of agreement.

On the benefits: Astrophysics-by-LLM is almost as good as human astrophysics in terms of feeding humanity's love of the subject;
it delivers benefits to physics (provided that it is asked to do so);
it can be used for military development;
it can contribute to remote sensing;
it creates new knowledge (provided that the projects it creates are indeed novel).
That is, astrophysics-by-LLM delivers many of the benefits that we have identified and associated with the practice of astrophysics.

So is everything fine? No everything is \emph{not fine}.
Recall (from our points of agreement in \secref{sec:facts}) that the practice of astrophysics cannot be learned from \emph{reading about astrophysics}, and that astrophysics is not about obtaining the answers, but rather the work we do, ourselves, to find those answers.
When we offload that work to LLMs, we are no longer doing astrophysics, we are no longer becoming astrophysicists, and, eventually, we no longer \emph{are} astrophysicists.
The let-them-cook policy, in the end, leads to the death of astrophysics, the end of astrophysics at universities, and the end of astrophysics education.
Astrophysics would no longer be \emph{by} humans, and then it would no longer be \emph{for} humans.

You might say: Perhaps let-them-cook is possible, but in a world in which \emph{both} humans and LLMs write, read, evaluate, and criticize papers, in a beautiful harmonic partnership.
Let's just coexist with the LLMs!
I love that, and I would love that solution.
However, there is no comparison of the rates at which LLMs and humans can produce papers.
If we take the ratio of minutes to months, or the ratio of the time for an LLM to completely do a scientific project to the same but for a (very capable) human (team), we get something like a factor of $10^5$.
LLM instances can be deployed on huge compute infrastructure, so the already-disturbing $10^5$ gets multiplied by the number of instances possible on available computing hardware, which is probably also in the $10^4$ range, in the astronomy-controlled computing capacity in the US alone, right now.
Of course some of the LLM papers will also require large compute to do their projects (astrophysics requires compute), but not all of them do.
With the LLMs (eventually) producing papers at $10^8$ or $10^9$ times the individual-human rate,\footnote{%
It is interesting to note, as is (effectively) noted in \cite{denario}, that even if the vast majority of LLM papers are technically or scientifically ``mid,'' at this rate, \emph{some} of the LLM-generated papers might be absolutely excellent and insightful.}
even the community of thousands of research-active human astrophysicists can't really meaningfully coexist with them.
Recall the points in \secref{sec:facts} that we are obliged to verify and cite the literature that is relevant to our work.

I'd be happy to hear how that argument is wrong.
It's probably wrong because there just aren't $10^9$ good project ideas arising each year.
So the LLMs would be ideas-limited, perhaps?
But the humans are not, and never will be, ideas-limited (every astrophysicist has more project ideas than time).
Thus even if the whole system is ideas-limited, the LLMs still overwhelmingly dominate astrophysics production.
This is related to the point above that we don't do astrophysics to \emph{find out the answers}:
In the let-them-cook scenario, all the humans can really do is read the outcomes of the astrophysics work being done by the far-more-productive LLMs.

There are other objections to let-them-cook, one of which is along the lines of the point of agreement about trust:
Right now the LLMs are not particularly trustworthy; in let-them-cook human astrophysicists would have to spend a lot of time checking or verifying LLM work.
LLMs might get more trustworthy with time, but it will likely be a while before we trust them as much as we trust code we write, say, with astropy \cite{astropy}, which is community built and vetted.

A comment to make about let-them-cook is that it would absolutely need to be associated with strong conventions or policies about transparency, disclosure, and reproducibility.
Transparency is paramount.
It is unconscionable to use an LLM to write words and not explain that; after all, the LLMs are, technically, plagiarists.\footnote{%
This point was made very strongly to me by Natalie B. Hogg \cite{natalie}, along with the point that we never know, with an LLM, what is the \emph{provenance} of something it writes.
Of course, since we all learn what we know and how to explain it by reading and talking to others, maybe even we humans are all plagiarists, and have inscrutable provenance also?
I don't think that is true, but an argument could be made.}
Transparency is required for the provenance point I discuss above in \secref{sec:facts}; it is critical to the integrity of the astrophysics literature.

But transparency is also required for \emph{reproducibility};
one very disturbing thing about the current LLMs is that they are not even close to reproducible.
Indeed they are designed to \emph{never} give the same answer to the same prompt.
Literally the opposite of reproducible, plus they are constantly upgraded.\footnote{%
The irreversible upgrading of systems is anathema to reproducibility.
For example the NASA \textsl{Horizons} system \url{https://ssd.jpl.nasa.gov/} that models the Solar System (for computing, for example, barycentric corrections for radial-velocity measurements and transit timing) is regularly upgraded without any preservation of past state, rendering almost every result in the observational study of exoplanets technically irreproducible.}
The reproducibility point, on its own, might be killing for let-them-cook, at least with the LLMs in their current forms.
Of course, as I commented in \secref{sec:facts}, there are many aspects of astrophysics that are not strictly reproducible, for very fundamental reasons.

By the way, if it really is true that the LLMs, in a matter of years, could increase the size of the astrophysics literature by factors of thousands or millions,
and if, in that scenario, almost all papers are written by LLMs and also read by LLMs,
then there is no reason to expect that the scientific literature would or should be written in human-readable languages.\footnote{%
A related but different point---about the computer languages in which LLMs will write code in the future---is being discussed in the vibe-coding arena \cite{bloom}.}
In this future, presumably ``reading the literature'' means having a translator LLM search and translate it for you.

\paragraph{Ban-and-punish:}
The other possible extreme policy proposal I could make (and, to be clear, I am also \emph{not} making) is that we \emph{fully outlaw} the LLMs, and do not permit anyone in astrophysics to use them for anything other than the most menial of tasks (like summarizing documentation), and we also enforce negative consequences for people or projects that use them.
Under this policy, nothing fundamental about the practice of astrophysics would different from what it was in 2019, say, with the exception that web searches for code issues, and code development environments, might be a bit better (or worse, or at least different).
Looking outward, the ban-and-punish attitude is (I believe, based on anecdotal evidence) most ascendant in the traditional humanities (for example, English, or history), and least ascendant in the professional schools (for example, museum studies, or management).

The one difference in the practice of astrophysics (relative to the practice in 2019, say), if the astrophysics community fully implemented the ban-and-punish policy, would be that we would have to spend significant energy, resources, and attention on \emph{policing our communities}.
After all, as the the LLMs get better, their work gets harder and harder to detect (or to distinguish from human activity), so if we, as a community, wanted to implement ban-and-punish, we would put ourselves into some kind of arms race---with the LLMs getting harder to detect and our detection technologies getting better---similar to the arms race escalating rapidly in the assignment and assessment of high school and university essays.\footnote{%
This arms race has been on for decades prior to the emergence of LLMs,
with lots of resources being applied to both sides of the arms race.
Interestingly, my institution NYU used for many years a company called \textsl{Turnitin}, which is the ``nice'' side of a company that also runs ``naughty'' websites that help students cheat.
So, just like in military arms races, real companies will benefit financially from playing both sides.}
It is obvious to me that this policing activity would end up becoming either impossible, or else would take a large fraction of our time and effort.

Ban-and-punish is an extreme policy, but it is consistent with almost all of our points of agreement (listed in \secref{sec:facts}), and it doesn't interfere with most of the key benefits of astrophysics (listed in \secref{sec:benefits}).
After all, it doesn't change the practice of astrophysics.
It just adds an onerous policing activity to our practice.
The only respect in which ban-and-punish does not conform to the points of agreement is that it is almost certainly not the most \emph{efficient} way to do astrophysics.
It is already the case that LLM systems can produce plausible scientific papers in minutes, and this capability will likely improve.
That's relevant to funding agencies, funding sources, and the success of early-career astrophysicists.
If a project or a person has a way to do their project faster and better, why would we stop them from using it?
That seems artificial and harmful, to our sources of funding, and to our people.

Perhaps even more importantly, \emph{ban-and-punish violates academic freedom}.
I am almost an academic-freedom absolutist; certainly I value very highly the fact that I can work on whatever I want, with whatever tools I want, with whoever I want, in whatever style I want.
It's not just a luxury:
Academic freedom makes possible the creativity and innovation that leads to discovery and advancement in astrophysics (and all fields).
Furthermore, in my opinion, academic freedom is not just the privilege of tenured faculty.
Everyone working in astrophysics, including every student, postdoc, junior faculty member, staff member, and engineer has---or really deserves---some degree of academic freedom.
We should operate our projects and our groups such that our people, no matter their rank, have as much freedom as possible.
That's just a good idea, but also it helps make our work consistent with the points of agreement about novelty, about people, and about efficiency.

To be very clear: I don't think ban-and-punish is good policy.
It artificially restricts what astrophysicists---and especially early-career astrophysicists---can do.
The ban-and-punish policy violates academic freedom, and it isn't consistent with our obligations to our funding sources and our public.
Plus, I am not a cop; ACAB.

\paragraph{Is there a middle way?}
There must be!
But I am not sure I can find it.
And even if I can find it, I don't see how we, as a community, could agree to it (and even less police or enforce it).
This isn't different from other ideas in scientific ethics (how can we agree on, police, and enforce ethical behavior?),
but it \emph{feels different} because the LLMs are qualitatively more capable and more available than other kinds of ethically questionable scientific channels, and because they are (currently) so untrustworthy.

One idea or rule of thumb is perhaps that we should or could interact with an LLM only in the same ways that we would interact with a non-coauthor colleague.\footnote{This suggestion is from Adrian Price-Whelan (Flatiron).}
You might ask your colleague for help finding something in the literature, but you wouldn't ask your non-coauthor colleague to \emph{write the introduction} of the paper you are writing.
You might ask your colleague to help speed up your code, but you wouldn't ask your colleague to \emph{write your code}.
Another idea is that, given our respect for our readers, we shouldn't ask them to engage with something that took way less time to write than to read.\footnote{This suggestion is from Joseph Long (Flatiron).}
I like these ideas; maybe they could evolve into conventions or community-supported best practices.

The only specific position I can put forward here is that transparency and disclosure are going to be important parts of whatever we do.
I think that's uncontroversial.
A hotter take is that the disclosure requirements we adopt might become quite onerous, since the LLMs are (by construction) not reproducible.
It might be that we must record and publish (in some form) every single complete LLM interaction we have in performing a scientific project.
This is a lot, but perhaps not dissimilar to the scientific notebook requirements associated with US National Institutes of Health grant awards \cite{nih}.

\paragraph{Comments:}
Here I make a final set of comments on the above, in no particular order.
The first comment I want to make is that it is hard for me to anticipate how what I've written here will be received.
I am not particularly a hater on the LLMs.
I use them in my research, and especially to learn about programming languages and environments with which I am unfamiliar.
For this very \documentname, I used LLMs to learn some details about \textsl{BibTeX} formatting, and also to get a summary of some ideas in animal rights (see above) and post-structuralist literary criticism (see below).

I said at the beginning that LLMs are not ``AIs.''
Because the LLMs are data-driven text-interpolators, I believe that their best and most useful role is in \emph{refactoring} or pivoting \emph{documentation}.
For example, when I used an LLM (\textsl{Claude} in this case) to learn about post-structuralism, I asked about strains of thought in literary criticism.
The LLM permitted me to do a kind of ``reverse look-up'' of that concept, starting at natural-language phrases like ``the idea that the meaning of a document is set by the reader, at the time of reading, not by the author, at the time of writing.''
Once that gave me the word ``post-structuralism,'' I worked through the Wikipedia page \cite{poststructuralism}, which is absolutely excellent.

\emph{Trust} is a critical idea among the points of agreement in \secref{sec:facts},
and trust is also one of the main issues that practitioners bring up when they use or interact with the LLMs.
How could we ever trust an LLM, especially when they, historically, can't tell you how many letter Rs there are in the word ``strawberry''?
As LLMs improve, trust will probably improve, both because the companies recognize this as a critical limitation, and because there are (I understand) large groups working on introspection, interpretability, explainability, and guardrails.
I am not sure we will ever trust an LLM as much as we trust the coordinates package, say, of \textsl{astropy} \cite{astropy},
but we might end up, on a timescale of months to years, considering LLM code and text to be substantially correct.
That would be great, but correctness is only one small part of trust.
A trusted partner is one that takes responsibility for their work.
The \textsl{astropy} codebase can't \emph{itself} take responsibility, but the specific authors of the code, and in particular the authors of the coordinates package, definitely \emph{can} take responsibility, and \emph{do} take responsibility.
They have careers and reputations and relationships on the line.
Entities like \textsl{Claude}---and its parent Anthropic---take no responsibility and never will.\footnote{%
This responsibility point is a big part of the reason we prefer open-source software projects over alternatives:
Open-source projects tend to have transparent authorship, bug-reporting systems, and responsive maintainers.
It's very hard to get a technical question or bug report answered by Apple or Microsoft.
This point implicitly brings up the question: What would be different if we had truly open-source LLMs, with open training and open weights?}

While above I said that citations should \emph{not} be seen as the ``coin of the realm,'' in fact they \emph{are} very materially relevant:
More highly published and more highly cited scholars, especially at early career stages, tend to fare better in job searches.
This makes sense because it is hard to assess the quality and impact of scientific work when that work is young and relatively untested; publication and citation counts serve as proxy.
Thus there will naturally be big pressures on scientists to use tools that speed analysis, writing, and publication.

Everything written here has been about Western academic astrophysics.
As I mentioned at the beginning, astronomy has been studied by countless civilizations, long before we established the Western traditions in which I currently work.
Non-Western and ancient astronomy practices did not adhere to many (maybe not even most) of the points of agreement in \secref{sec:facts}.
Also, for many civilizations, astronomy \emph{did} have important clinical value:
Astronomical observations were used to develop navigation technologies (especially over oceans \cite{navigation}),
to deliver timing for planting and harvesting operations,
and to record in reproducible form various calendar events, like birth months.\footnote{%
Astronomers like to hate on astrology, but astrological signs (when they are properly maintained in terms of the Sun's actual angular location among the constellations) can be seen as a coordinate-free, universally accessible, reproducible record of the time of year, divorced from state power.}
Astronomy provides a universally visible set of clocks that create and support opportunities across a wide range of human activities.

I have made the strong---and possibly unrealistic---assumption that, over the next months and years, the LLMs will get \emph{far better}.
Although people are using them to design, execute, and write up projects, in fact most of what the LLMs write is not currently excellent; it requires a lot of checking and revision and discarding (see, for example, \cite{bloomvibe}).
Right now, the LLMs make only \emph{slop}, and a present-day scenario in which a high-performance deployment of LLMs generating $10^8$ submittable papers per year would produce horrifying nightmare of useless-to-damaging trash.
Some of the generated papers would be novel, correct, and interesting perhaps, but it would be essentially impossible to find them, even armed with additional LLMs.
There is every possibility that this ``LLM moment'' (some would say ``GenAI moment'') is a fad or a flash and we will look with pity upon any astronomer who spent time writing a 20-page \documentname{} about them at the beginning of 2026.

In putting together this \documentname, one objection I heard to the idea that LLMs could write papers is that human input is required to give documents \emph{meaning};\footnote{Phil Marshall (Stanford), private communication.}
perhaps, for this reason, nothing a machine generates could have meaning?
I find this point appealing: Aren't we, authors, imbuing our words with meaning through our essential humanity?
But as appealing as this is, I believe this to be wrong.
I am on the side of the post-structuralists:
The meaning of a document is set---or almost entirely set---by the reader, or at the time of reading \cite{deadauthor}.
Obviously this is debatable, and perhaps the rise of the LLMs will lead to a reconsideration of meaning.
Perhaps meaning involves that which is transmitted from the author's knowledge to the reader's knowledge.
If the LLMs don't ``know'' anything, then, in this view, the documents they ``write'' might end up being meaningless.

Finally: Why did I write this \documentname?
I wrote this because I became concerned about some of the ideas circulating in the astrophysics community about LLMs and their capabilities, conflating what are (in my view) text-interpolators with what are (in my view) scientists.
There is, obviously, a very wide range of opinions, within the responsible astrophysics community, about what LLMs are and how they can be used as tools (see \cite{ting} and \cite{trotta} for two very different but responsible perspectives).
Ultimately, I think the real question we face---if we do indeed face a question---is not the question of \emph{how} we do astrophysics.
It is the question of \emph{why} we do astrophysics.

\bibliography{why}

@article{denario,
  title={The {Denario} project: Deep knowledge {AI} agents for scientific discovery},
  author={Villaescusa-Navarro, Francisco and others},
  journal={arXiv preprint arXiv:2510.26887},
  year={2025}
}

@article{gaia,
  title={The {Gaia} {M}ission},
  author={Gaia~Collaboration},
  journal={Astronomy \& Astrophysics},
  volume={595},
  pages={A1},
  year={2016},
  publisher={EDP sciences}
}

@article{dpac,
  title={Gaia: Organisation and challenges for the data processing},
  author={Mignard, Fran{\c{c}}ois and others},
  journal={Proceedings of the International Astronomical Union},
  volume={3},
  number={S248},
  pages={224--230},
  year={2007},
  publisher={Cambridge University Press}
}

@article{jwst,
  title={The {James Webb Space Telescope}},
  author={Gardner, Jonathan P. and others},
  journal={Space Science Reviews},
  volume={123},
  number={4},
  pages={485--606},
  year={2006},
  publisher={Springer}
}

@article{lsst,
  title={{LSST}: From science drivers to reference design and anticipated data products},
  author={Ivezi{\'c}, {\v{Z}}eljko and others},
  journal={The Astrophysical Journal},
  volume={873},
  number={2},
  pages={111},
  year={2019},
  publisher={IOP Publishing}
}

@article{galah,
  title={The GALAH survey: scientific motivation},
  author={De Silva, G. M. and others},
  journal={Monthly Notices of the Royal Astronomical Society},
  volume={449},
  number={3},
  pages={2604--2617},
  year={2015},
  publisher={Oxford University Press}
}

@article{monitor,
  title={A Photometricity and extinction monitor at the {Apache Point Observatory}},
  author={Hogg, David W. and Finkbeiner, Douglas P. and Schlegel, David J. and Gunn, James E.},
  journal={The Astronomical Journal},
  volume={122},
  number={4},
  pages={2129},
  year={2001},
  publisher={IOP Publishing}
}

@article{ubercal,
  title={An improved photometric calibration of the {Sloan Digital Sky Survey} imaging data},
  author={Padmanabhan, Nikhil and others},
  journal={The Astrophysical Journal},
  volume={674},
  number={2},
  pages={1217},
  year={2008},
  publisher={IOP Publishing}
}

@article{astrometrynet,
  title={Astrometry.net: {Blind} astrometric calibration of arbitrary astronomical images},
  author={Lang, Dustin and Hogg, David W. and Mierle, Keir and Blanton, Michael and Roweis, Sam},
  journal={The Astronomical Journal},
  volume={139},
  number={5},
  pages={1782},
  year={2010},
  publisher={IOP Publishing}
}

@article{opendata,
  title={Ten simple rules for the care and feeding of scientific data},
  author={Goodman, Alyssa and others},
  journal={PLoS Computational Biology},
  volume={10},
  number={4},
  pages={e1003542},
  year={2014},
  publisher={Public Library of Science San Francisco, USA}
}

@article{emcee,
  title={{emcee}: The {MCMC} hammer},
  author={Foreman-Mackey, Daniel and Hogg, David W. and Lang, Dustin and Goodman, Jonathan},
  journal={Publications of the Astronomical Society of the Pacific},
  volume={125},
  number={925},
  pages={306},
  year={2013},
  publisher={IOP Publishing}
}

@article{vera,
  title={Dark matter in spiral galaxies},
  author={Rubin, Vera C.},
  journal={Scientific American},
  volume={248},
  number={6},
  pages={96--109},
  year={1983},
  publisher={JSTOR}
}

@article{causal,
  title={Can large language models infer causation from correlation?},
  author={Jin, Zhijing and Liu, Jiarui and Lyu, Zhiheng and Poff, Spencer and Sachan, Mrinmaya and Mihalcea, Rada and Diab, Mona and Sch{\"o}lkopf, Bernhard},
  journal={arXiv preprint arXiv:2306.05836},
  year={2023}
}

@article{sdss,
  title={The {Sloan Digital Sky Survey}: Technical summary},
  author={York, Donald G. and others},
  journal={The Astronomical Journal},
  volume={120},
  number={3},
  pages={1579},
  year={2000},
  publisher={IOP Publishing}
}

@article{sdssiii,
  title={{SDSS-III}: Massive spectroscopic surveys of the distant {Universe}, the {Milky Way}, and extra-solar planetary systems},
  author={Eisenstein, Daniel J. and others},
  journal={The Astronomical Journal},
  volume={142},
  number={3},
  pages={72},
  year={2011},
  publisher={IOP Publishing}
}

@article{sdssiv,
  title={{Sloan Digital Sky Survey IV}: Mapping the {Milky Way}, nearby galaxies, and the distant {Universe}},
  author={Blanton, Michael R. and others},
  journal={The Astronomical Journal},
  volume={154},
  number={1},
  pages={28},
  year={2017},
  publisher={IOP Publishing}
}

@article{sdssv,
  title={{Sloan Digital Sky Survey V}. {P}ioneering Panoptic Spectroscopy},
  author={Kollmeier, Juna A. and others},
  journal={The Astronomical Journal},
  volume={171},
  number={1},
  pages={52},
  year={2025},
  publisher={IOP Publishing}
}

@article{keplerresults,
  title={Planetary candidates observed by {Kepler}. {VIII}. {A} fully automated catalog with measured completeness and reliability based on {Data Release 25}},
  author={Thompson, Susan E. and others},
  journal={The Astrophysical Journal Supplement Series},
  volume={235},
  number={2},
  pages={38},
  year={2018},
  publisher={IOP Publishing}
}

@article{lightecho,
  title={Light echoes from ancient supernovae in the {Large Magellanic Cloud}},
  author={Rest, Armin and others},
  journal={Nature},
  volume={438},
  number={7071},
  pages={1132--1134},
  year={2005},
  publisher={Nature Publishing Group UK London}
}

@article{singletransit,
  title={The population of long-period transiting exoplanets},
  author={Foreman-Mackey, Daniel and Morton, Timothy D. and Hogg, David W. and Agol, Eric and Sch{\"o}lkopf, Bernhard},
  journal={The Astronomical Journal},
  volume={152},
  number={6},
  pages={206},
  year={2016},
  publisher={IOP Publishing}
}

@phdthesis{phdthesis,
  title={On the evolution of field galaxies},
  author={Hogg, David W.},
  year={1998},
  school={California Institute of Technology}
}

@article{oti,
  title={Orbital torus imaging: Using element abundances to map orbits and mass in the {Milky Way}},
  author={Price-Whelan, Adrian M. and others},
  journal={The Astrophysical Journal},
  volume={910},
  number={1},
  pages={17},
  year={2021},
  publisher={IOP Publishing}
}

@article{age,
  title={{Planck 2018 results. VI.} {Cosmological} parameters},
  author={Aghanim, N. and others},
  journal={Astronomy \& Astrophysics},
  volume={641},
  pages={A6},
  year={2020}
}

@article{adelberger,
  title={New tests of {Einstein's} equivalence principle and {Newton's} inverse-square law},
  author={Adelberger, E. G.},
  journal={Classical and Quantum Gravity},
  volume={18},
  number={13},
  pages={2397},
  year={2001},
  publisher={IOP Publishing}
}

@article{gravityprobeb,
  title={{Gravity probe B}: Final results of a space experiment to test general relativity},
  author={Everitt, C. W. Francis and others},
  journal={Physical Review Letters},
  volume={106},
  number={22},
  pages={221101},
  year={2011},
  publisher={APS}
}

@book{oceanobserving,
  title={Discovering the Ocean from Space: The Unique Applications of Satellite Oceanography},
  author={Robinson, Ian S.},
  year={2010},
  publisher={Springer Science \& Business Media}
}

@book{oppy,
  title={Atom and Void: Essays on Science and Community},
  author={Oppenheimer, J Robert},
  year={2014},
  publisher={Princeton University Press}
}

@book{kant,
  author = {Kant, Immanuel},
  title = {Groundwork of the Metaphysics of Morals},
  year = {1785},
  translator = {Gregor, Mary J.},
  publisher = {Cambridge University Press},
  address = {Cambridge},
  series = {Cambridge Texts in the History of Philosophy},
  edition = {Revised},
  origyear = {1785},
  origpublisher = {Johann Friedrich Hartknoch},
  note = {Trans. Mary J. Gregor. Originally published as \textsl{Grundlegung zur Metaphysik der Sitten}}
}

@book{mill,
  author    = {Mill, John Stuart},
  title     = {Utilitarianism},
  year      = {1863},
  publisher = {Parker, Son, and Bourn},
  address   = {London}
}

@book{simons,
  title={The Man Who Solved the Market: How Jim Simons Launched the Quant Revolution},
  author={Gregory Zuckerman},
  year={2019},
  publisher={Portfolio}
}

@article{curiosity,
  title={The physicists philosophy of physics},
  author={Peebles, P. James E.},
  journal={arXiv preprint arXiv:2401.16506},
  year={2024}
}

@article{imperative,
  title={The imperative to reduce carbon emissions in astronomy},
  author={Stevens, Adam R. H. and Bellstedt, Sabine and Elahi, Pascal J. and Murphy, Michael T.},
  journal={Nature Astronomy},
  volume={4},
  number={9},
  pages={843--851},
  year={2020},
  publisher={Nature Publishing Group UK London}
}

@article{low-carbon,
  title={Astronomy in a low-carbon future},
  author={Matzner, Christopher D. and others},
  journal={arXiv preprint arXiv:1910.01272},
  year={2019}
}

@article{effectivealtruism,
  title={The definition of effective altruism},
  author={MacAskill, William},
  journal={Effective altruism: Philosophical issues},
  volume={2016},
  number={7},
  pages={10},
  year={2019},
  publisher={Oxford University Press Oxford}
}

@incollection{deadauthor,
  author    = {Barthes, Roland},
  title     = {The Death of the Author},
  booktitle = {Image-Music-Text},
  editor    = {Heath, Stephen},
  translator = {Heath, Stephen},
  publisher = {Hill and Wang},
  address   = {New York},
  year      = {1977},
  pages     = {142--148},
  note = {Originally published in 1967 as \textsl{La mort de l'auteur}}
}

@article{paperslop,
  title={Quantifying large language model usage in scientific papers},
  author={Liang, Weixin and others},
  journal={Nature Human Behaviour},
  pages={1--11},
  year={2025},
  publisher={Nature Publishing Group UK London}
}

@article{water,
  title={The water use of data center workloads: A review and assessment of key determinants},
  author={Lei, Nuoa and Lu, Jun and Shehabi, Arman and Masanet, Eric},
  journal={Resources, Conservation and Recycling},
  volume={219},
  pages={108310},
  year={2025},
  publisher={Elsevier}
}

@book{animallib,
  author    = {Singer, Peter},
  title     = {Animal Liberation: A New Ethics for Our Treatment of Animals},
  publisher = {New York Review/Random House},
  year      = {1975},
  address   = {New York}
}

@book{marssux,
  title={A City on Mars: Can We Settle Space, Should We Settle Space, and Have We Really Thought This Through?},
  author={Weinersmith, Kelly and Weinersmith, Zach},
  year={2025},
  publisher={Penguin Group}
}

@article{astropy,
  title={The astropy project: Building an open-science project and status of the v2.0 core package},
  author={The Astropy Collaboration and others},
  journal={The Astronomical Journal},
  volume={156},
  number={3},
  pages={123},
  year={2018},
  publisher={IOP Publishing}
}

@book{navigation,
  title={Hawaiki Rising: H{\=o}k{\=u}le{\textquotesingle}a, Nainoa Thompson, and the Hawaiian Renaissance},
  author={Low, Sam},
  year={2013},
  publisher={Island Heritage Publishing},
  address={Waipahu, HI}
}

@article{undetectable,
  title={Measuring the undetectable: Proper motions and parallaxes of very faint sources},
  author={Lang, Dustin and Hogg, David W. and Jester, Sebastian and Rix, Hans-Walter},
  journal={The Astronomical Journal},
  volume={137},
  number={5},
  pages={4400},
  year={2009},
  publisher={IOP Publishing}
}

@article{chatauthor,
  title={Rapamycin in the context of Pascal’s Wager: generative pre-trained transformer perspective},
  author={Transformer, ChatGPT Generative Pre-trained and Zhavoronkov, Alex},
  journal={Oncoscience},
  volume={9},
  pages={82},
  year={2022}
}

@article{ting,
  title={Artificial intelligence compels the astronomy community to rethink research identity and redefine excellence},
  author={Ting, Yuan-Sen},
  journal={Nature Astronomy},
  volume = 9,
  pages= {317-318},
  year={2025},
  publisher={Nature Publishing Group UK London}
}

@article{trotta,
  title={The indiscriminate adoption of {AI} threatens the foundations of academia},
  author={Trotta, Roberto},
  journal={Nature Astronomy},
  volume = 9,
  pages = {1748–1749},
  year={2025},
  publisher={Nature Publishing Group UK London}
}

@article{conference,
  title={{AI} bots wrote and reviewed all papers at this conference},
  author={Gibney, Elizabeth},
  journal={Nature},
  volume={646},
  number={8086},
  pages={786--786},
  year={2025},
  publisher={Nature}
}

@article{arithmetic,
  title={Training Verifiers to Solve Math Word Problems},
  author={Cobbe, Karl and others},
  journal={arXiv preprint arXiv:2110.14168},
  year={2021}
}

@article{turing,
  title={Computing Machinery and Intelligence},
  author={Turing, Alan M.},
  journal={Mind},
  volume={59},
  number={236},
  pages={433--460},
  year={1950},
  publisher={Oxford University Press}
}

@misc{nih,
  author = {National Institutes of Health},
  title = {{Intramural Electronic Lab Notebook Policy}},
  note = {\url{https://oir.nih.gov/sourcebook/intramural-program-oversight/electronic-lab-notebooks/intramural-electronic-lab-notebook-policy} accessed 2026 February 1.}
}

@misc{icmje,
  author = {International Committee of Medical Journal Editors},
  title = {{V. Use} of Artificial Intelligence in Publishing},
  note = {\url{https://www.icmje.org/recommendations/browse/artificial-intelligence/} accessed 2026 February 5.},
}

@misc{cope,
  author = {Committee on Publication Ethics (COPE) Council},
  title = {{COPE} position -- {Authorship and AI} -- {English}},
  note = {\url{https://doi.org/10.24318/cCVRZBms} accessed 2026 February 5.}
}

@misc{natalie,
  author = {Natalie B. Hogg},
  title = {On the use of {ChatGPT} in academia},
  note = {\url{https://nataliebhogg.com/2024/01/15/claiming-the-right-to-be-unhappy/} accessed 2026 February 5.}
}

@misc{bloom,
  author = {Joshua Bloom},
  title = {The Last Programming Languages Created by People},
  note = {\url{https://medium.com/@profjsb/the-last-programming-languages-created-by-people-6aef63e310b7} accessed 2026 February 5.}
}

@misc{bloomvibe,
  author = {Joshua Bloom},
  title = {({Mis})Adventures of {GenAI} in the Scientific Workflow},
  note = {\url{https://medium.com/@profjsb/mis-adventures-of-genai-in-the-scientific-workflow-d2ff1d804850} accessed 2026 February 6.}
}

@misc{poststructuralism,
  author = {Wikipedia},
  title = {Post-structuralism},
  note = {\url{https://en.wikipedia.org/wiki/Post-structuralism} accessed 2026 February 5.}
}

@book{dontlook,
  title = {Don't Look Up},
  year = 2021,
  publisher = {Netflix},
  note = {Dir. Adam McKay}
}

\end{document}